\documentclass{article}
\usepackage{spconf,amsmath,graphicx}

\title{Multitrack Music Transformer}
\name{Hao-Wen Dong\thanks{Contact: \texttt{\href{mailto:hwdong@ucsd.edu}{hwdong@ucsd.edu}}} \hspace{1em} Ke Chen \hspace{1em} Shlomo Dubnov \hspace{1em} Julian McAuley \hspace{1em} Taylor Berg-Kirkpatrick \vspace{-1.25ex}}
\address{University of California San Diego}
\usepackage[T1]{fontenc} % add special characters (e.g., umlaute)
\usepackage[utf8]{inputenc} % set utf-8 as default input encoding
\usepackage[hidelinks,pagebackref,colorlinks=true]{hyperref}
\usepackage{url}
\usepackage{color}

\usepackage{cite}
\usepackage{inconsolata}
\usepackage{amsmath}
\usepackage{amssymb}
\usepackage{booktabs}
\usepackage{microtype}
\usepackage{enumitem}
\usepackage[noabbrev,capitalise]{cleveref}
\usepackage{bm}
\usepackage{bbm}
\usepackage{multirow}
\usepackage{tabularx}

\setlist{parsep=0ex,topsep=0.5ex,itemsep=0ex,leftmargin=2em}
\crefformat{footnote}{#2\footnotemark[#1]#3}
\crefrangeformat{footnote}{#3\footnotemark[#1]#4--#5\footnotemark[#2]#6}
\crefmultiformat{footnote}{#2\footnotemark[#1]#3}{\textsuperscript{,}#2\footnotemark[#1]#3}{\textsuperscript{,}#2\footnotemark[#1]#3}{\textsuperscript{,}#2\footnotemark[#1]#3}
\graphicspath{{figs/}}

\newcolumntype{C}{>{\centering\arraybackslash}X}
\definecolor{darkred}{rgb}{0.6,0.0,0.0}
\definecolor{darkblue}{rgb}{0.0,0.0,0.75}
\definecolor{myblue}{rgb}{0,0.3,0.6}
\hypersetup{linkcolor=myblue,urlcolor=myblue,citecolor=myblue,anchorcolor=myblue}

\begin{document}
\ninept
\maketitle
\begin{abstract}
Existing approaches for generating multitrack music with transformer models have been limited in terms of the number of instruments, the length of the music segments and slow inference. This is partly due to the memory requirements of the lengthy input sequences necessitated by existing representations. In this work, we propose a new multitrack music representation that allows a diverse set of instruments while keeping a short sequence length. Our proposed Multitrack Music Transformer (MMT) achieves comparable performance with state-of-the-art systems, landing in between two recently proposed models in a subjective listening test, while achieving substantial speedups and memory reductions over both, making the method attractive for real time improvisation or near real time creative applications. Further, we propose a new measure for analyzing musical self-attention and show that the trained model attends more to notes that form a consonant interval with the current note and to notes that are 4N beats away from the current step.
\end{abstract}

\setlength{\abovedisplayskip}{1ex}
\setlength{\belowdisplayskip}{1ex}
\setlength{\abovedisplayshortskip}{1ex}
\setlength{\belowdisplayshortskip}{1ex}

%=====================
\section{Introduction}
%=====================
\label{sec:introduction}

Prior work has investigated various approaches for symbolic music generation \cite{briot2017survey,ji2020survey}, among which, the transformer model \cite{vaswani2017transformer} has become popular given its recent successes in piano music generation \cite{huang2019musictransformer,huang2020remi,hsiao2021transformer,muhamed2021transformergan}. At the core of a transformer model is the self-attention mechanism that allows the model to dynamically attend to different parts of the input sequence and aggregate information from the whole sequence. Such capabilities make it suitable for modeling the complex structures and textures in music. However, while prior work has also explored applying transformer models to generate multitrack music \cite{musenet,donahue2019lakhnes,ens2020mmm,vonrutte2022figaro}, successful implementations have only been reported either on a limited set of instruments \cite{musenet,donahue2019lakhnes} or short music segments \cite{ens2020mmm,vonrutte2022figaro}. This is partly due to the long sequence length in existing multitrack music representations, which results in a large memory requirement in training. For example, a GPU with 11GB of memory can only generate 29 seconds of music on average using the REMI+ representation \cite{vonrutte2022figaro} on an orchestral music dataset. Moreover, it can only generate less than four notes per second. These limitations together pose a challenge in scaling transformer models to longer music with many instruments, e.g., orchestral music, and for real-time use cases, e.g., automatic improvisation and human-AI music co-creation.

In this paper, we propose a new multitrack music representation to address the long sequence issue in existing multitrack music representations. Using the proposed representation, we present the Multitrack Music Transformer (MMT) for multitrack music generation. Unlike a standard transformer model, the proposed model uses a decoder-only transformer with multi-dimensional inputs and outputs to reduce its memory complexity. On an orchestral dataset, we show that our proposed model can generate longer music in a faster inference speed than two existing approaches. Through a subjective listening test, we show that the proposed model achieves reasonably good performance in terms of coherence, richness and arrangement as well as the overall quality. Moreover, our proposed representation allows a trained autoregressive model to generate music for a specific set of instruments, a task that has not been well studied in prior work.

Further, while the transformer model has been widely used on symbolic music, it remains unclear how self-attention work for symbolic music. Understanding musical self-attention could reveal future research directions in improving transformer models for music. To the best of our knowledge, existing analysis \cite{huang2019musictransformer,huang2018visualizing,chen2021transformer,ziyu2021musebert} provides only case studies on few selected samples, lacking a systematic analysis on self-attention for music. Hence, we propose a new quantity to measure the average attention weights that a transformer model assigns to a certain key of a certain difference from the query. Our analysis shows that the proposed model learns a relative self-attention for certain aspects of music, specifically, beat, position and pitch.

Our proposed model provides a novel foundation for future work exploring longer-form and real-time capable multitrack music generation. The systematic analysis also provide insights into improving the self-attention mechanism for music. Audio samples can be found on our demo website.\footnote{\url{https://salu133445.github.io/mmt/}\label{fn:demo}} For reproducibility, all source code, hyperparameters and pretrained models are available at \url{https://github.com/salu133445/mmt}.

\begin{table}
    \footnotesize
    \centering
    \vspace{-1.5ex}
    \caption{Comparisons of related transformer-based music models.}
    \label{tab:comparison}
    \vspace{1ex}
    \begin{tabularx}{\linewidth}{lCCCC}
        \toprule
        \multirow{1}{*}{Model} &Multitrack &Instrument control &Compound tokens &Generative modeling\\
        \midrule
        REMI~\cite{huang2020remi} &&&&\checkmark\\
        MMM~\cite{ens2020mmm} &\checkmark &&&\checkmark\\
        CP~\cite{hsiao2021transformer} &&&\checkmark &\checkmark\\
        MusicBERT~\cite{zeng2021musicbert} &\checkmark &&\checkmark\\
        FIGARO~\cite{vonrutte2022figaro} &\checkmark &&&\checkmark\\
        \cmidrule(lr){1-5}
        MMT (ours) &\checkmark &\checkmark &\checkmark &\checkmark\\
        \bottomrule
    \end{tabularx}
    \vspace{-0.5ex}
\end{table}

\begin{figure*}
    \footnotesize
    \centering
    \begin{minipage}{0.74\linewidth}
    \includegraphics[width=\linewidth]{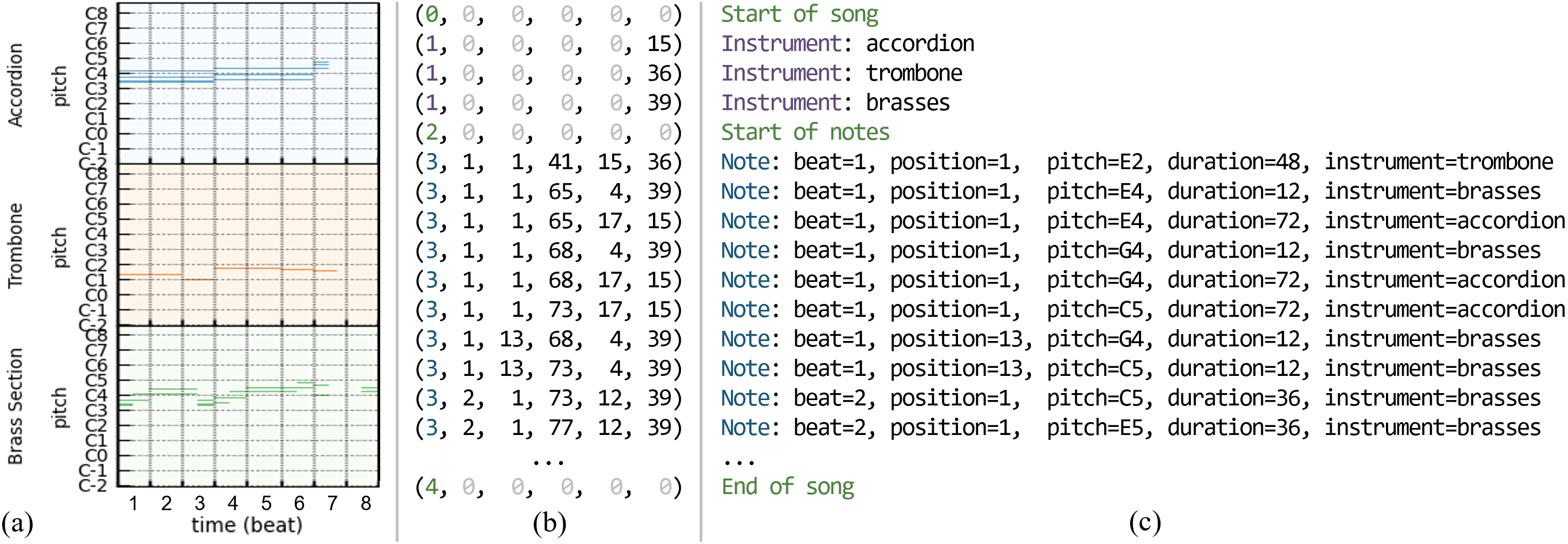}
    \end{minipage}
    \hfill
    \begin{minipage}{0.21\linewidth}\vspace{-2ex}
    \caption{An example of the proposed representation---(a) an example of the first eight beats of a song in the orchestra dataset, shown as a multitrack piano roll, (b) the same song encoded by our proposed representation, where the grayed out zeros denote undefined values and (c) a human-readable translation of the codes shown in (b).}
    \label{fig:representation}
    \end{minipage}\vspace{-1ex}
\end{figure*}

%=====================
\section{Related Work}
%=====================

\textbf{Multitrack music generation.}~~Prior work has explored various approaches for symbolic music generation~\cite{briot2017survey,ji2020survey}, among which generating multitrack music is considered more challenging for its complex interdependency between voices and instruments. In~\cite{dong2018musegan,dong2018binarymusegan}, the authors used a convolutional generative adversarial network to generate short, five-track pop music segments. In~\cite{simon2018multitrackmusicvae}, the authors used a variational autoencoder with recurrent neural networks to learn a latent space for multitrack measures. In~\cite{donahue2019lakhnes,musenet}, the authors used decoder-only transformer models to generate four-track game music and multi-instrument classical music, respectively. In~\cite{vonrutte2022figaro}, the authors used a transformer model to generate multitrack music given a fine-grained description of the characteristics of the desired music. Unlike these systems, our proposed model is built upon a more compact representation that allows it to accommodate longer sequences under the same GPU memory constraint.

\vspace{1ex}\noindent\textbf{Transformers for symbolic music.}~~Another relevant line of research is on modeling symbolic music with transformer models \cite{vaswani2017transformer}. Some prior work focused on unconditioned generation, including generating piano music \cite{huang2020remi,hsiao2021transformer}, lead sheets \cite{wu2020jazz,wu2023compose}, guitar tabs \cite{chen2020guitar} and multitrack music \cite{donahue2019lakhnes,musenet} from scratch. Others studied controllable music generation \cite{shih2022themetransformer,vonrutte2022figaro}, music style transfer \cite{wu2023musemorphose}, polyphonic music score infilling \cite{chang2021infiliing} and general-purpose pretraining for symbolic music understanding \cite{chou2021midibert,zeng2021musicbert,ziyu2021musebert}. In this work, we focus on unconditioned generation for evaluation purposes. However, our proposed model can also generate music for a set of instruments specified by the user.

%========================
\section{Proposed Method}
%========================
\label{sec:method}

%-------------------------------
\subsection{Data Representation}
%-------------------------------
\label{sec:representation}

We represent a music piece as a sequence of events $\mathbf{x} = (\mathbf{x}_1, \dots, \mathbf{x}_n)$, where each event $\mathbf{x}_i$ is encoded as a tuple of six variables: $$(x_i^\mathit{type}, x_i^\mathit{beat}, x_i^\mathit{position}, x_i^\mathit{pitch}, x_i^\mathit{duration}, x_i^\mathit{instrument})\,.$$ The first variable $x^\mathit{type}$ determines the type of the event, among the following five event types:
\begin{itemize}
    \item \textit{Start-of-song}: Indicates the beginning of the song.
    \item \textit{Instrument}: Specifies an instrument used in the song.
    \item \textit{Start-of-notes}: Indicates the end of the instrument list and the beginning of the note list. (This event splits the sequence into two parts: a list of instrument events followed by a list of note events, making a trained autoregressive model readily applicable to instrument-informed generation task; see \cref{sec:model}.)
    \item \textit{Note}: Specifies a note, whose onset, pitch, duration and instrument are defined by the other five variables: $x^\mathit{beat}$, $x^\mathit{position}$, $x^\mathit{pitch}$, $x^\mathit{duration}$ and $x^\mathit{instrument}$.
    \item \textit{End-of-song}: Indicates the end of the song.
\end{itemize}
For any non-note-type event, the variables $x^\mathit{beat}$, $x^\mathit{position}$, $x^\mathit{pitch}$, $x^\mathit{duration}$, $x^\mathit{instrument}$ are set to zero, which is reserved for undefined values. \cref{fig:representation} shows an example of our proposed representation.

Following \cite{huang2020remi}, we decompose the note onset information into  beat and position information, where $x^\mathit{beat}$ denotes the index of the beat that the note lies in, and $x^\mathit{position}$ the position of the note within that beat. To be specific, the actual onset of the note is equivalent to $r\cdot x^\mathit{beat} + x^\mathit{position}$, where $r$ is the temporal resolution of a beat. For simplicity, we assume that the beats are always a quarter note apart in this work. This decomposition reduces the size of the vocabulary and helps the model learn the music meter system, as evidenced by \cite{huang2020remi}. For the duration field, following \cite{vonrutte2022figaro}, we only allow a carefully-chosen set of common note duration values and replace any duration outside of this set with the closest known duration. For the instrument field, we map similar MIDI programs to the same instrument to reduce the total number of instruments, resulting in 64 unique instruments from the 128 MIDI programs. For example, both `acoustic grand piano' and `bright acoustic piano' are mapped to the same `piano' instrument. Note that the reduced number of MIDI instruments do not affect the encoded sequence length, and the majority of savings comes from combining multiple variables into a tuple.

We note that the proposed representation leads to a significantly shorter sequence length as compared to two existing representations \cite{ens2020mmm,vonrutte2022figaro} for multitrack music generation. On an orchestral dataset \cite{crestel2017lop}, an encoded sequence of length 1,024 using our proposed representation can represent 2.6 and 3.5 times longer music samples compared to \cite{ens2020mmm} and \cite{vonrutte2022figaro}, respectively. Further, because the timing information is embedded into each note event, the proposed representation is invariant to permutation, i.e., reordering the note events do not affect the decoded music. For the sake of autoregressive training for the transformer model, we sort the notes with respect to the beat field, and subsequently the position, pitch, duration, instrument fields. This allows a trained autoregressive model to be readily applicable to the song continuation task.

%-----------------
\subsection{Model}
%-----------------
\label{sec:model}

\begin{figure}
    \small
    \centering
    \includegraphics[width=.83\linewidth]{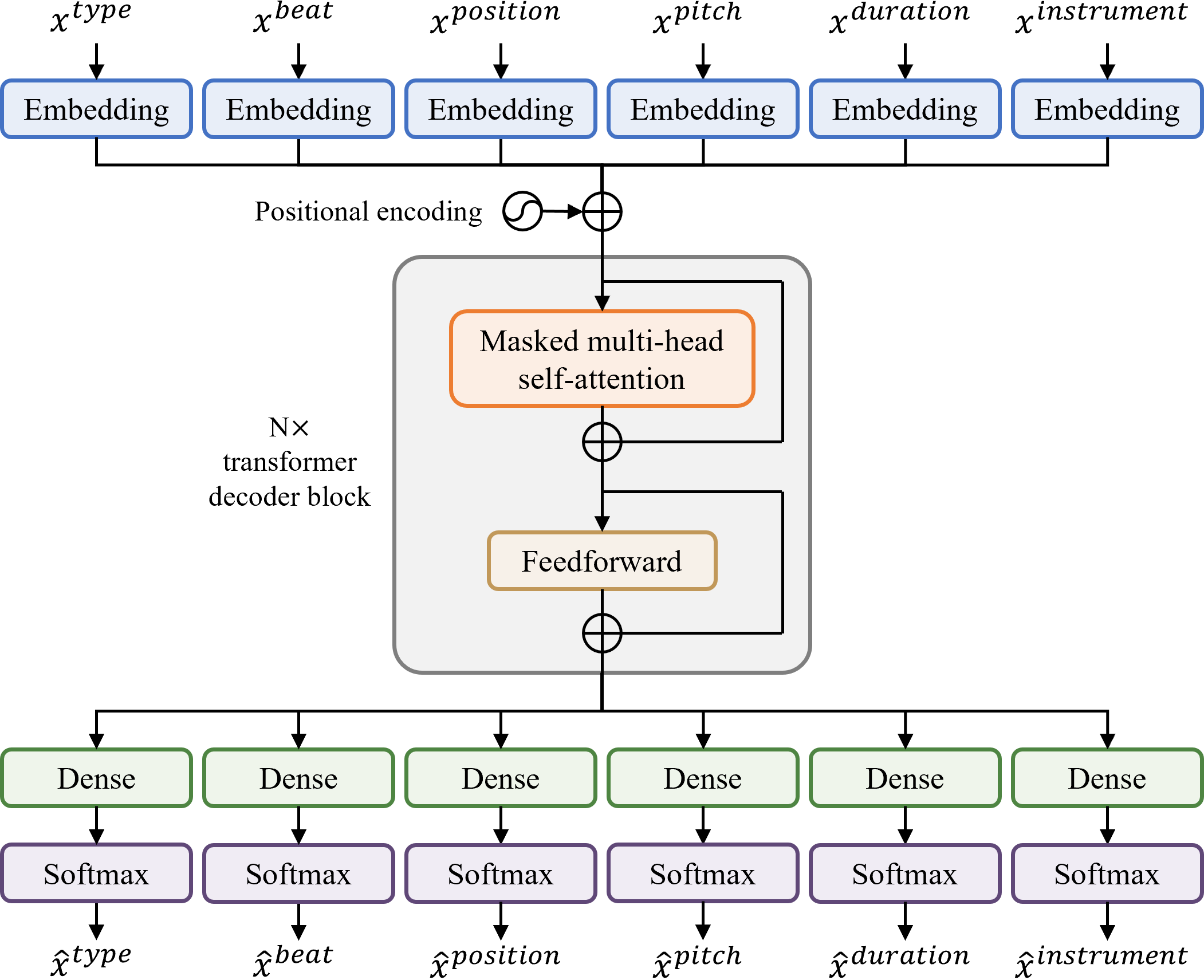}
    \caption{Illustration of the proposed MMT model.}
    \label{fig:model}
\end{figure}

We present the Multitrack Music Transformer (MMT) for generating multitrack music using the representation proposed in~\cref{sec:representation}. We base the proposed model on a decoder-only transformer model~\cite{liu2018transformerdecoder,brown2020gpt3}. Unlike a standard transformer model, whose inputs and outputs are one-dimensional, the proposed model has multi-dimensional input and output spaces similar to~\cite{hsiao2021transformer}, as illustrated in \cref{fig:model}. The model is trained to minimize the sum of the cross entropy losses of different fields under an autoregressive setting. We adopt a learnable absolute positional embedding~\cite{vaswani2017transformer}. Once the training is done, the trained transformer model can be used in three different modes, depending on the inputs given to the model to start the generation:
\begin{itemize}
    \item \textbf{Unconditioned generation}: Only a `start-of-song' event is provided to the model. The model generates the instrument list and subsequently the note sequence.
    \item \textbf{Instrument-informed generation}: The model is given a `start-of-song' event followed by a sequence of instrument codes and a `start-of-notes' event to start with. The model then generates the note sequence. Note that we need the `start-of-notes' event as it marks the end of the instrument list, otherwise the model may continue to generate instrument events.
    \item \textbf{$N$-beat continuation}: All instrument and note events in the first $N$ beats are provided to the model. The model then generates subsequent note events that continue the input music.
\end{itemize}

During inference, the sampling process is stopped when an `end-of-song' event is generated or the maximum sequence length is reached. We adopt the top-$k$ sampling strategy on each field and set $k$ to 10\% of the number of possible outcomes per field. Moreover, since the type and beat fields in our representation are always sorted, we further enforce a monotonic constraint during decoding. For example, when sampling for $x_{i + 1}^\mathit{type}$, we set the probability of getting a value smaller than $x_{i}^\mathit{type}$ to zero. This prohibits the model from generating events in certain invalid order, e.g., an `note' event before an `instrument' event.

Finally, while existing multitrack music generation systems \cite{ens2020mmm,vonrutte2022figaro} need to combine several generated tokens to form a note, the proposed MMT model generates a note at each inference step, i.e., a line in \cref{fig:representation}(b) and (c). This offers MMT a significantly faster inference speed and smaller memory footprint thanks to the reduced size of the self-attention matrix. However, since MMT predicts the six output fields nonautoregressively (i.e., independently), it cannot model the interdependencies between these fields of the same note. We will discuss this trade-off between time/memory complexity and modeling capacity in \cref{sec:subjective}.

%================
\section{Results}
%================
\label{sec:result}

%----------------------------
\subsection{Experiment Setup}
%----------------------------

In this work, we consider the Symbolic Orchestral Database (SOD)~\cite{crestel2017lop}. We set the temporal resolution to 12 time steps per quarter note. We discard tempo and velocity information as not all data contains such information. Further, we discard all drum tracks. We end up with 5,743 songs (357 hours). We reserve 10\% of the data for validation and 10\% for testing. We use MusPy~\cite{dong2020muspy} to process the data. For the proposed MMT model, we use 6 transformer decoder blocks, with a model dimension of 512 and 8 self-attention heads. All input embeddings have 512 dimensions. We trim the code sequences to a maximum length of 1,024 and a maximum beat of 256. During training, we augment the data by randomly shifting all the pitches by $s \sim U(-5, 6)\ (s \in \mathbb{Z})$ semitones and randomly selecting a starting beat. We validate the model every 1K steps and stop the training at 200K steps or when there was no improvements for 20 validation rounds. We render all audio samples using FluidSynth with the MuseScore General SoundFont. We encourage the readers to listen to the sample generated music on our demo website.\cref{fn:demo} % for the unconditional generation, instrument-informed generation and $N$-beat continuation tasks

\begin{table*}
    \footnotesize
    \centering
    \vspace{-1.5ex}
    \caption{Performance comparison of our proposed model against the baseline models. Mean values and 95\% confidence intervals are reported.}
    \label{tab:performance}
    \vspace{1ex}
    \begin{tabular}{lccccccc}
        \toprule
        &\multirow{2}{*}[-.5ex]{\shortstack{Number of\\parameters}} &\multirow{2}{*}[-.5ex]{\shortstack{Average sample\\length (sec)}} &\multirow{2}{*}[-.5ex]{\shortstack{Inference speed\\(notes per second)}}&\multicolumn{4}{c}{Subjective listening test results}\\
        \cmidrule(lr){5-8}
        &&& &Coherence &Richness &Arrangement &Overall\\
        \midrule
        MMM~\cite{ens2020mmm} &19.81 M &\underline{38.69} &\underline{5.66} &3.48 $\pm$ 0.35 &3.05 $\pm$ 0.38 &3.28 $\pm$ 0.37 &3.17 $\pm$ 0.43\\
        REMI+~\cite{vonrutte2022figaro} &20.72 M &28.69 &3.58 &\textbf{3.90 $\pm$ 0.52} &\textbf{3.74 $\pm$ 0.21} &\textbf{3.74 $\pm$ 0.44} &\textbf{3.77 $\pm$ 0.41}\\
        MMT (ours) &19.94 M &\textbf{100.42} &\textbf{11.79} &\underline{3.55 $\pm$ 0.46} &\underline{3.53 $\pm$ 0.35} &\underline{3.40 $\pm$ 0.44} &\underline{3.33 $\pm$ 0.47}\\
        \bottomrule
    \end{tabular}
\end{table*}

%-------------------------------------
\subsection{Subjective Listening Test}
%-------------------------------------
\label{sec:subjective}

To assess the quality of music samples generated by our proposed model, we conducted a listening test with 9 music amateurs recruited from our social networks, where all survey participants can play at least one musical instrument. In the questionnaire, each participant was asked to listen to 10 audio samples generated by each model and rate each audio sample according to three criteria---\textit{coherence}, \textit{richness} and \textit{arrangement}.\footnote{To be specific, we ask the following questions: \textit{coherence}---``Is it temporally coherent? Is the rhythm steady? Are there many out-of-context notes?''; \textit{richness}---``Is it rich and diverse in musical textures? Are there any repetitions and variations? Is it too boring?''; \textit{arrangement}---``Are the instruments used reasonably? Are the instruments arranged properly?''} We compared the MMT model against two baseline models based on the standard decoder-only transformer model. The first baseline model used the MultiTrack representation proposed in the MMM model \cite{ens2020mmm}, where we replaced the bar tokens with beat tokens. The other used a simplified version of the REMI+ representation used in the FIGARO model \cite{vonrutte2022figaro}, where we removed the time signature, tempo and chord tokens as such information is not generally available in our datasets. We will refer to the two baseline models as the MMM and REMI+ models. For a fair comparison, we trimmed all generated samples to a maximum of 64 beats. Moreover, as discussed in \cref{sec:introduction}, the long sequence length of existing multitrack music representations restricts the model from learning long-term dependencies. Hence, we also computed the mean length of the generated samples and the inference speed in this experiment.

We summarize in \cref{tab:performance} the evaluation results. Compared to the MMM model, our proposed MMT model achieves a higher score across all criteria. Further, MMT generates 2.6 times longer samples and is twice faster in inference speed. As compared to the REMI+ model, our proposed model achieves a mean opinion score (MOS) of 3.33, while the REMI+ model achieves an MOS of 3.77. However, MMT can generate 3.5 times longer samples and is 3.3 times faster in inference speed. This is because the baseline models need multiple inference passes to combine several generated tokens and form a note, whereas the MMT model generate a note in a single inference pass. Finally, we note that while offering a faster inference speed and longer generated sample length, our proposed model cannot model the interdependencies between the six output heads as it predicts each field independently. For example, the REMI+ model first generates an instrument token and then generates the pitch token given the instrument token, which allows the model to rule out unsuitable pitches for that particular instrument. In contrast, the MMT model samples from each output head independently. We can clearly observe this trade-off between quality and between time/memory complexity can be clearly observed from \cref{tab:performance}.

%--------------------------------
\subsection{Objective Evaluation}
%--------------------------------
\label{sec:objective}

In addition the subjective listening test, we follow \cite{mogren2016crnngan,wu2020jazz} and measure the pitch class entropy, scale consistency and groove consistency for evaluating the performance of the proposed model on the unconditioned generation task. For these metrics, we consider a closer value to that of the ground truth better. \cref{tab:objective} shows the evaluation results. We can see that the REMI+ model achieves closest values to those of the ground truth. We also notice that while the MMM model result in closer values of pitch class entropy and scale consistency to those of the ground truth, it achieves a lower score in the subjective listening test presented in \cref{sec:subjective} than our proposed MMT model.

\begin{table}
    \footnotesize
    \centering
    \vspace{-1.5ex}
    \caption{Objective evaluation results. Mean values and 95\% confidence intervals are reported. A closer value to that of the ground truth is considered better.}
    \label{tab:objective}
    \vspace{1.5ex}
    \begin{tabularx}{\linewidth}{lCCC}
        \toprule
        &Pitch class entropy &Scale consistency (\%) &Groove consistency (\%)\\
        \midrule
        Ground truth &2.974 $\pm$ 0.018 &92.26 $\pm$ 1.25 &93.05 $\pm$ 1.00\\
        \cmidrule(lr){1-4}
        MMM~\cite{ens2020mmm} &\underline{2.884 $\pm$ 0.023} &\underline{93.13 $\pm$ 0.49} &91.90 $\pm$ 0.64\\
        REMI+~\cite{vonrutte2022figaro} &\textbf{2.897 $\pm$ 0.019} &\textbf{93.12 $\pm$ 0.51} &\textbf{92.90 $\pm$ 0.49}\\
        MMT (ours) &2.802 $\pm$ 0.025 &94.74 $\pm$ 0.42 &\underline{92.09 $\pm$ 0.49}\\
        \bottomrule
    \end{tabularx}
\end{table}

%----------------------------------
\subsection{Musical Self-attention}
%----------------------------------
\label{sec:attention}

Despite the growing interests in applying transformer models to music, little effort has been made to understand how self-attention works for symbolic music---existing analyses~\cite{huang2019musictransformer,huang2018visualizing,chen2021transformer,ziyu2021musebert} provide only case studies on few selected samples. In this section, we aim to investigate musical self-attention in a systematic way. To this end, we propose two new quantities to measure the average relative attention. Mathematically, given a test set $\mathcal{D}$, we define the \textit{mean relative attention} for a field $d$ (e.g., pitch or beat) as:
\begin{equation}
    \gamma_k^{(d)} = \frac{\sum_{\mathbf{x} \in \mathcal{D}} \sum_{s > t}\,a_{s, t}(\mathbf{x})\;\mathbbm{1}_{x^{(d)}_t - x^{(d)}_s = k}}{\sum_{\mathbf{x} \in \mathcal{D}} \sum_{s > t}\,a_{s, t}(\mathbf{x})}\,,
\end{equation}
where $\mathbbm{1}[\cdot]$ is the indicator function and $a_{s, t}(\mathbf{x}) \in [0, 1]$ denotes the attention weight assigned by $\mathbf{x}_s$ to $\mathbf{x}_t$. Intuitively, $\gamma_k^{(d)}$ measures the average attention weight that model assigns to a certain key of a certain difference from the query. Note that each attention head has its own attention weight $a_{s, t}$ and thus its own $\gamma_k$. Moreover, we notice that $\gamma_k^{(d)}$ is biased towards differences that occur more frequently. Thus we further propose the \textit{mean relative attention gain}:
\begin{equation}
    \tilde{\gamma}_k^{(d)} = \gamma_k^{(d)} - \frac{\sum_{\mathbf{x} \in \mathcal{D}} \sum_{s > t}\,\mathbbm{1}_{x^{(d)}_t - x^{(d)}_s = k}}{\sum_{\mathbf{x} \in \mathcal{D}} \sum_{s > t}\,1}\,,
\end{equation}
which measures the difference between $\gamma_k^{(d)}$ and the same quantity obtained by assuming a uniform attention matrix.

\begin{figure}
    \newlength{\temp}
    \setlength{\temp}{2.5cm}
    \footnotesize
    \hspace*{1.61cm}$\star$\hspace{.52cm}$\star$\hspace{.52cm}$\star$\hspace{.52cm}$\star$\hspace{.52cm}$\star$\hspace{.52cm}$\star$\hspace{.52cm}$\star$\hspace{.52cm}$\star$\hspace{.52cm}$\star$\\
    \includegraphics[width=\linewidth,height=\temp,trim={0 0 0 2.5mm},clip]{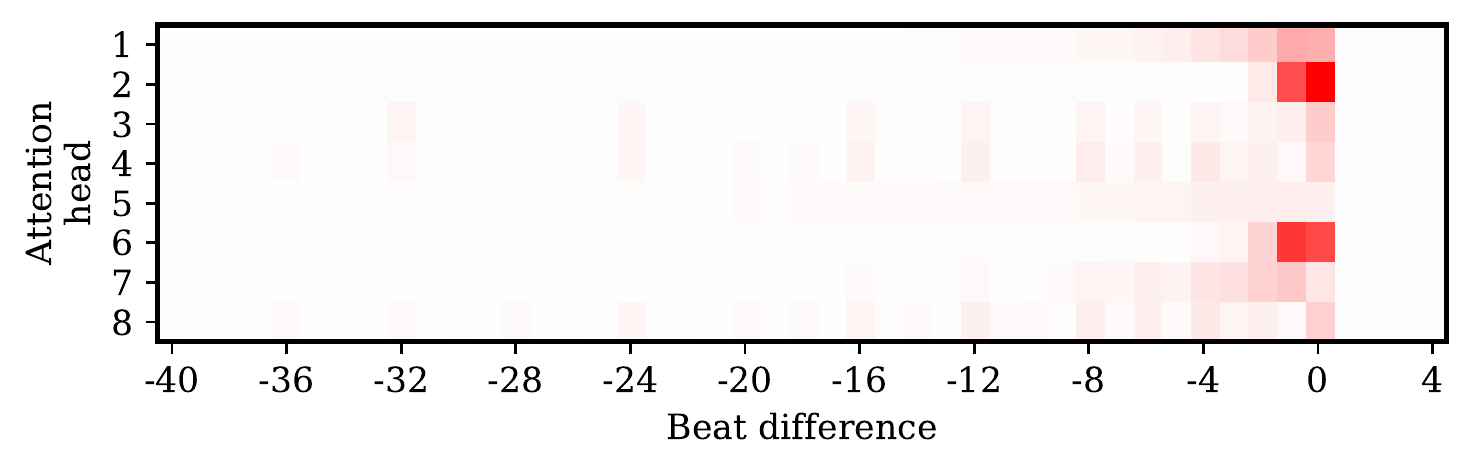}\\[-4ex]
    (a)\\[1.5ex]
    \hspace*{2.7cm}{\tiny$_\nearrow$~\scriptsize 8th note behind}{\tiny$_\nearrow$~\scriptsize same position}\hspace{.2cm}{\tiny$_\nearrow$~\scriptsize 8th note ahead}\\[-.5ex]
    \includegraphics[width=\linewidth,height=\temp,trim={0 0 0 2.5mm},clip]{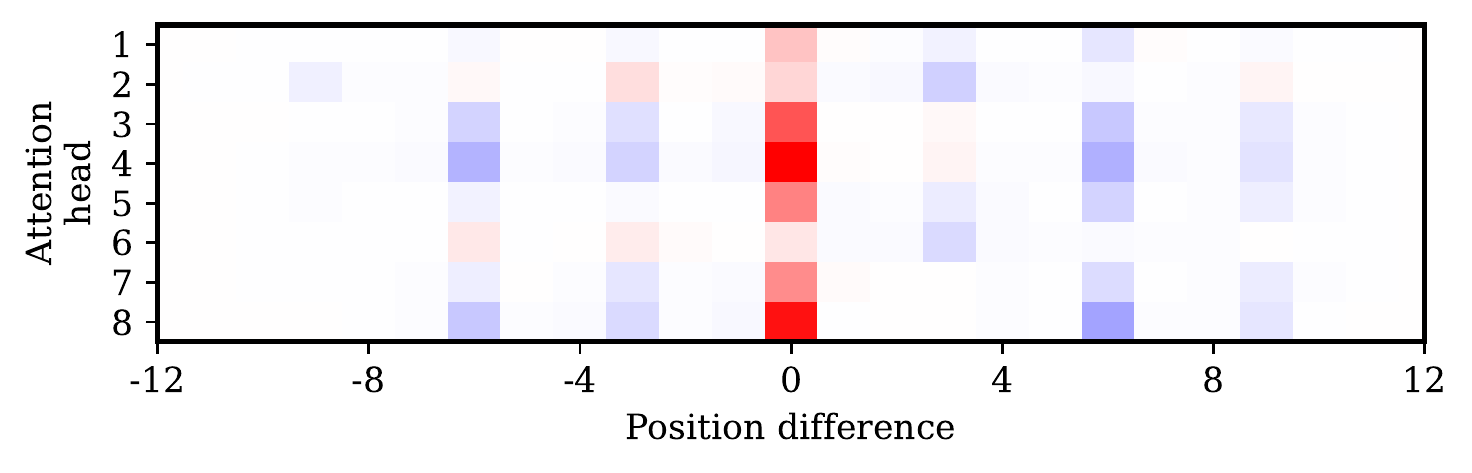}\\[-4ex]
    (b)\\[1.5ex]
    \hspace*{3.42cm}{\scriptsize same pitch\tiny$_\nwarrow$}\hspace{.17cm}{\scriptsize 4th\tiny$_\nwarrow$}\hspace{.19cm}{\tiny$_\nearrow$\scriptsize 5th}\hspace{.17cm}{\tiny$_\nearrow$\scriptsize 8th (octave)}\\[-.5ex]
    \includegraphics[width=\linewidth,height=\temp,trim={0 0 0 2.5mm},clip]{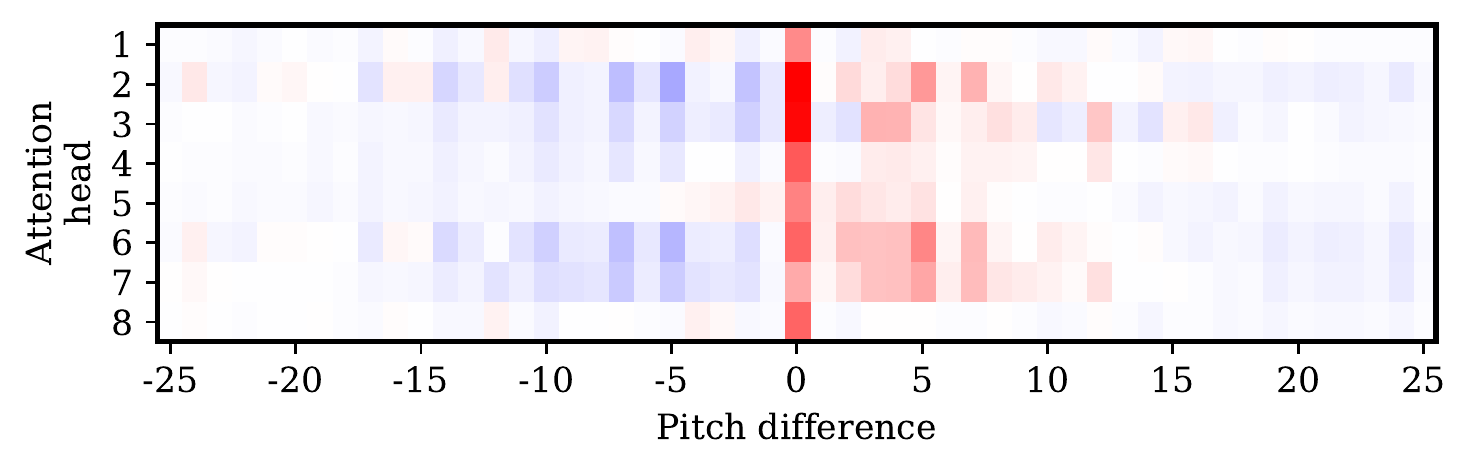}\\[-4ex]
    (c)
    \caption{Mean relative attention gains (a) $\tilde{\gamma}_k^\mathit{beat}$, (b) $\tilde{\gamma}_k^\mathit{position}$ and (c) $\tilde{\gamma}_k^\mathit{pitch}$ (see \cref{sec:attention} for definitions) of a trained MMT model. Red and blue colors indicate \textcolor{darkred}{positive} and \textcolor{darkblue}{negative} values, respectively.}
    \label{fig:attention}
\end{figure}

In this experiment, we compute $\tilde{\gamma}_k^\mathit{beat}$, $\tilde{\gamma}_k^\mathit{position}$ and $\tilde{\gamma}_k^\mathit{pitch}$ on 100 test samples for the last attention layer of a trained MMT model. As shown in \cref{fig:attention}(a), we can see that the 2nd and 6th attention heads attend more to nearby beats, while the other attention heads attend to beats in further past. In addition, several attention heads assign relatively larger weights to the beats that are $4N$ (i.e., 4, 8, 12, 16, etc.) beats away from the current one, as highlighted by the `$\star$' symbols. From \cref{fig:attention}(b) we observe that the model pays most attention to notes that have the same position as the current note. That is, a note on beat attends more to the last note on beat, and a note off beat attends more to the last note off beat. \cref{fig:attention}(c) shows that the model attends more to pitches within one octave above, and it pays more attention to pitches that form a consonant interval with the current note, e.g., a 4th, a 5th and an octave. We note that the learned self-attention generally comply with music theory principles.

While recent advances in symbolic music generation has borrowed various techniques from natural language modeling, music is fundamentally different from text in that music has a underlying temporal axis embedded and contains strong recurrence patterns in many aspects. Our analysis here shows that our proposed model learns a relative self-attention for certain aspects of music, specifically, beat, position and pitch. We hope our analysis can shed light on further improvements in optimizing the self-attention mechanism for symbolic music modeling.

%===================
\section{Conclusion}
%===================

We have presented the Multitrack Music Transformer for multitrack music generation. Built upon a new multitrack representation, our proposed model can generate longer multitrack music in a faster inference speed than two existing approaches. We showed in a subjective listening test that the proposed model perform reasonably well against the two baseline models in terms of the quality of the generated music. Through a systematic analysis, we showed that our proposed model learns relative self-attention in certain aspects of music such as beats, positions and pitches. Our findings provide a novel foundation for future work exploring longer-form, real-time capable multitrack music generation and improving the self-attention mechanism for music.

%=========================
\section{Acknowledgements}
%=========================

Hao-Wen thanks J. Yang and Family Foundation and Taiwan Ministry of Education for supporting his PhD study. This project has received funding from the European Research Council (ERC REACH) under the European Union’s Horizon 2020 research and innovation programme (Grant agreement \#883313).

% \vfill\pagebreak

\bibliographystyle{IEEEbib}
\bibliography{ref}

\end{document}